\documentclass[seceq]{ptptex}

\usepackage{graphicx}

\def\Vect#1{\mbox{\boldmath $#1$}}
\def\cl{{\cal L}}
\def\dgam{\dot{\gamma}}
\def\cP{{\cal P}}
\def\cQ{{\cal Q}}
\title{
Mode-coupling theory of sheared dense granular liquids%
}

\author{
Hisao \textsc{Hayakawa}
and Michio \textsc{Otsuki}%
}

\inst{
Yukawa Institute for Theoretical Physics, Kyoto University,  Kitashirakawaoiwake-cho, Sakyo,
Kyoto 606-8502, JAPAN
}

\abst{
Mode-coupling theory (MCT) of sheared dense granular liquids 
is formulated. 
Starting from the Liouville equation of granular particles, the generalized Langevin equation is derived with
the aid of the projection operator technique. The MCT equation for the density correlation function
obtained from the generalized Langevin equation is almost equivalent to
MCT equation for elastic particles under the shear.
It is found that there should be the plateau in the density correlation function.
}

\begin{document}

\maketitle

\section{Introduction}
The rheology of granular materials is one of central concerns in granular physics. 
Although the granular particles exhibit unusual behaviors\cite{jaeger96},
 Liu and Nagel\cite{lin98} suggested the possibility that the jamming transition of granular particles under the shear 
may be regarded as the glass transition at zero temperature. 
Since then  the jamming has been recognized as one of the most important concepts in
nonequilibrium rheology\cite{miguel,cates98,coniglio00,silbert02,makse02,fierro04,coniglio05,hatano07a,hatano07b,majumar07,kurchan07}.
However, so far there exist few theoretical works which discuss direct connections  
between the jamming transition of dense granular materials and those of the other materials 
such as colloidal suspensions\cite{miyazaki02,fuchs02,miyazaki04,fuchs05}
from microscopic point of view.

The lack of the direct evidences of the universality is due to the difficulties in describing the dynamics of 
dense granular media.
 On the other hand, indirect evidences of universality of the jamming transitions are being accumulated with time; 
(i) There are strong similarities for the emergence of dynamical yield stress in sheared granular materials\cite{hatano07a} 
to that of colloidal suspensions.\cite{fuchs05}
(ii) The steep peak of the nonlinear susceptibility so called  $\chi_4$ is observed  in the vicinity of the jamming transition
in experiments of shaken granular materials\cite{douchat07}.
(iii) The long-tails in nearly elastic sheared granular fluids are almost the same as those for a system
 of elastic particles with a thermostat.\cite{otsuki07} 

We recognize that the gas kinetic theory can successfully describe the behaviors of  relatively dilute granular flows 
including correlation effects\cite{otsuki07,brilliantov,jenkins,garzo,lutsko05,saitoh,hh-mo07}, 
but  it is obvious that we cannot use the kinetic theory in describing the jamming transition.
On the other hand, many of theoretical works\cite{coniglio00,makse02,fierro04,coniglio05} 
starting from statics of granular particles
focus on the introduction of the effective 
temperature based on 
the idea of Edwards' compactivity in thermodynamics of granular media.\cite{edwards}
Thus, so far there are few liquid theories in characterizing dense granular liquids.
It should be noted that we do not have to introduce any exotic temperature such as compactivity
 in describing sheared granular liquids, because it is standard to use
the granular temperature defined by $T\equiv \sum_i \langle m(\Vect{v}_i-\Vect{u})^2\rangle/Nd$ where $m$, $\Vect{v}_i$, $\Vect{u}$, $N$ and $d$
are respectively the mass of each particle, the velocity of $i-$th particle, the velocity of the flow field, the number of particles and the dimension.
In addition, it is known that the steady uniform shear flow has the uniform granular temperature except for the boundary layer\cite{hatano07a,hatano07b} 
in sheared dense granular systems, which is in contrast
to the case of the relatively dilute sheared granular flows driven by the boundary.\cite{saitoh}
Therefore, it may be possible to construct a liquid theory or the mode-coupling theory (MCT) 
by using the granular temperature to characterize the liquids near the jamming transition.  

In this paper we derive MCT equation for dense granular liquids 
from the Liouville
equation of granular fluids.
The derivation is formal and analogous to that for the ideal glass transition.\cite{goetz,binder,RC05} 
The organization of this paper is as follows. In the next section, we introduce the Liouville equation for granular gases.
In section 3, we formally derive the generalized Langevin equation by using the technique of the projection operator.
In section 4, we apply the generalized Langevin equation for the field variables such as the density field under the shear.
In section 5, we will derive the generalized Langevin equation for the density correlation function which is not a closed
equation because of the formal representation of the memory kernel. In section 6, we introduce the mode-coupling
approximation to obtain a closed equation for the density correlation function. The result is similar to that for 
ideal glass transition. In section 7, we discuss what we can predict from MCT  and future problems.  We also summarize our results.
In Appendix A, we show the details of the derivation of the generalized Langevin equation. In Appendix B, we summarize 
the expression of the correlation functions in the presence of the shear. In Appendix C, we discuss the derivation of MCT for
the sheared dense granular liquids.

\section{Liouville equation}

Let us consider a system of $N$ identical hard spherical and smooth particles with their diameters $\sigma$ and the constant of the restitution $e$ less than 
unity in $d-$dimensional space.
The basic equation to describe the dynamics of the classical particles is  the Liouville equation.\cite{zwanzig,hansen}  
The Liouville equation for granular fluids
has been discussed by Brey {\it et al.}\cite{brey97,dufty06,dufty06b} some times ago.
Let $i\cl_{tot}$ be the total Liouvillian which operates  any physical function $A(\Gamma(t))$ as
\begin{equation}\label{Liouville}
\frac{dA(\Gamma(t))}{d{t}}=i\cl_{tot} A(\Gamma(t)) , \quad A(\Gamma(t))=e^{i\cl_{tot} t}A(\Gamma(0)), 
\end{equation}
where $\Gamma(t)$ is the phase variable.  The total Liouvillian consists of three parts, the free part, the collision part and the shear part. 
We denote $i\cl_{tot}$ as $i\cl_{tot}=i\cl+i\cl_s$, where
$i\cl$ is  the sum of the free part and the collision part 
\begin{equation}\label{cl}
i\cl=\sum_{j=1}^N\Vect{v}_j\cdot\frac{\partial}{\partial \Vect{r}_j}+\frac{1}{2}\sum_j\sum_{k\ne j}T_{jk}
\end{equation}
with the velocity $\Vect{v}_j$ of $j-$th particle and the collision operator
\begin{equation}\label{collision}
T_{jk}=\sigma^{d-1}\int d\hat{\Vect{\sigma}}\delta(\Vect{r}_{jk}-\Vect{\sigma})
\Theta(-\Vect{g}_{jk}\cdot\hat{\Vect{\sigma}})|\hat{\Vect{\sigma}}\cdot\Vect{g}_{jk}|
(b_{jk}-1),
\end{equation}
where $\Vect{r}_{jk}=\Vect{r}_j-\Vect{r}_k$, $\Vect{g}_{jk}=\Vect{v}_j-\Vect{v}_k$, $\Theta(x)=1$ for $x>0$ and $\Theta(x)=0$ for otherwise.
$\hat{\Vect{\sigma}}$ is the unit normal vector at contact and $\Vect{\sigma}=\sigma\hat{\Vect{\sigma}}$. Here $b_{jk}$ is the collision operator acting on any function
$X(\Vect{v}_j,\Vect{v}_k)$ as
\begin{equation}\label{b_jk}
b_{jk}X(\Vect{v}_j,\Vect{v}_k)=X(b_{jk}\Vect{v}_j,b_{jk}\Vect{v}_k)=X(\Vect{v}_j',\Vect{v}_k')
\end{equation}
where the precollisional velocities $\Vect{v}_j$  and $\Vect{v}_k$  change into the postcollisional velocities 
$\Vect{v}_j'$ and $\Vect{v}_k'$, respectively.
When $b_{jk}$ acts on $\Vect{g}_{jk}$, $\Vect{g}_{jk}$ changes as
\begin{equation}
b_{jk}\Vect{g}_{jk}=\Vect{g}_{jk}-(1+e)(\hat{\Vect{\sigma}}\cdot\Vect{g}_{jk})\hat{\Vect{\sigma}} .
\end{equation}
The Liouville operator to represent the shear flow $i\cl_s$ is
\begin{equation}\label{L_s}
i\cl_s=\dgam\sum_{j=1}^N\left(y_j\frac{\partial}{\partial x_j}-v_{y,j}\frac{\partial}{\partial v_{x,j}}\right) ,
\end{equation}
where the macroscopic velocity under the shear flow is assumed to be
\begin{equation}
u_{\beta}=\dgam y \delta_{\beta,x},
\end{equation}
which recovers the equation of motion under the shear flow.\cite{evans}
We also note that eq.(\ref{L_s}) can be derived from the DOLLS Hamiltonian.\cite{evans,hoover}

It should be noted that the Liouvillian is not self-adjoint 
because of the violation of time reversal symmetry for each collision. 
The adjoint Liouvillian is defined for the equation of the phase function or $N-$body distribution function 
$\rho(\Gamma(t))$
\begin{equation}\label{rho}
\rho(\Gamma(t))=
e^{-i\bar{\cl}_{tot} t}\rho(\Gamma(0)), \quad \frac{d \rho(\Gamma(t))}{d t}=-i\bar{\cl}_{tot}\rho(\Gamma(t)).
\end{equation}
The average of a physical quantity is defined as
\begin{equation}\label{average}
\langle A(t)\rangle \equiv \int d\Gamma \rho(\Gamma)A(\Gamma(t))=\int d\Gamma A(\Gamma)\rho(\Gamma(t)) .
\end{equation}
From (\ref{Liouville}), (\ref{rho}) and (\ref{average}) we obtain the following relations
\begin{equation}\label{average2}
\int d\Gamma \rho(\Gamma)e^{i\cl_{tot} t}A(\Gamma)=\int d\Gamma A(\Gamma)e^{-i\bar\cl_{tot} t}\rho(\Gamma)
\end{equation}
and
\begin{equation}\label{average3}
\int d\Gamma \rho(\Gamma)i\cl_{tot} A(\Gamma)=-\int d\Gamma A(\Gamma) i\bar\cl_{tot} \rho(\Gamma) .
\end{equation}
The adjoint Liouvillian is obtained from the integration by parts of eq.(\ref{average3})\cite{dufty06}
\begin{equation}
i\bar{\cl}_{tot}=\sum_{j=1}^N\Vect{v}_j\cdot
\frac{\partial}{\partial \Vect{r}_j}+\frac{1}{2}\sum_j\sum_{k\ne j}\bar{T}_{jk}+i\bar\cl_s ,
\end{equation}
where
\begin{equation}
\bar{T}_{jk}=\sigma^{d-1}\int d\hat{\Vect{\sigma}}\Theta(\hat{\Vect{\sigma}}\cdot\Vect{g}_{jk})|\hat{\Vect{\sigma}}\cdot\Vect{g}_{jk}|[e^{-2}\delta(\Vect{r}_{jk}-\Vect{\sigma})b_{jk}^{-1}+
\delta(\Vect{r}_{jk}+\Vect{\sigma})]
\end{equation}
 and 
\begin{equation}
i\bar{\cl}_s=\dgam \sum_{j=1}^N\left\{\frac{\partial}{\partial x_j} y_j-\frac{\partial}{\partial v_{x,j}}v_{y,j}\right\} .
\end{equation}
Here $b_{jk}^{-1}$ is the inverse operator of $b_{jk}$ which satisfies $b_{jk}^{-1}b_{jk}=b_{jk}b_{jk}^{-1}=1$ and
\begin{eqnarray}
b_{jk}^{-1}\Vect{g}_{jk}&=&\Vect{g}_{jk}-\frac{1+e}{e}(\hat{\Vect{\sigma}}\cdot\Vect{g}_{jk})\hat{\Vect{\sigma}} . 
\end{eqnarray}

Before closing this section, we add some remarks.
First, since $i\cl_{tot}$ is not self-adjoint, the evloution operator acting on  $A^*(\Gamma(t))$ which is the complex conjugate of $A(\Gamma(t))$
 is neither  $-i\cl_{tot}$  nor 
$-i\bar{\cl}_{tot}$ in (\ref{rho}).
Second, the collision operators $T_{ij}$ and $\bar{T}_{ij}$ are reduced to the known Liouvillian for hard core particles in the limit of $e=1$.\cite{Resibois}
Third, the collision operators are exact for inelastic hard spherical granular particles for any density.

\section{Generalized Langevin equation}

What we are interested in is the time correlation function between $A(t)\equiv A(\Gamma(t))$ and $B$, the abbreviation of $B(\Gamma(0))$, as
\begin{eqnarray}\label{III-1}
C_{AB}(t)&\equiv& \langle A(t) B^*\rangle \nonumber \\
&=&\int d\Gamma [e^{i\cl_{tot} t} A(\Gamma)]B^*(\Gamma)\rho(\Gamma)
=\int d\Gamma A(\Gamma)e^{-i\bar{\cl}_{tot} t}[\rho(\Gamma)B^*(\Gamma)] ,
\end{eqnarray}
where 
we use the definition of $i\bar{\cl}_{tot}$ for the last expression.
The final expression of eq.(\ref{III-1}) is obtained from the integration by parts with the assumption
of zero net currents through the boundaries.

It is not easy to handle eq.(\ref{III-1}), because the adjoint Liouvillian $i\bar{\cl}_{tot}$ has the property\cite{dufty06b} 
\begin{equation}\label{dufty-A27}
i\bar{\cl}_{tot}[\rho(\Gamma) B^*(\Gamma)]
=i\bar{\cl}_{tot}[\rho(\Gamma)]B^*(\Gamma)+\rho(\Gamma) i\cl^-_{tot}B^*(\Gamma), 
\end{equation}
where
\begin{equation}\label{dufty-A32}
i\cl^-_{tot}\equiv \sum_{j=1}^N \Vect{v}_j\cdot\frac{\partial}{\partial \Vect{r}_j}-\frac{1}{2}\sum_j\sum_{k\ne j} \bar{T}^{-}_{jk}+i\bar{\cl}_s
\end{equation}
with
\begin{equation}\label{dufty-A33}
T^{-}_{jk}\equiv \sigma^{d-1}\int d\hat{\Vect{\sigma}}
\Theta(\hat{\Vect{\sigma}}\cdot \Vect{g}_{jk})|\hat{\Vect{\sigma}}\cdot\Vect{g}_{jk}|\delta(\Vect{r}_{jk}-\Vect{\sigma})(b^{-1}_{jk}-1).
\end{equation}
However, when the initial phase function $\rho(\Gamma)$ is invariant in the time evolution, {\it i.e.} 
\begin{equation}\label{steady-cd}
\bar{\cl}_{tot}\rho(\Gamma)=0,
\end{equation}
the simple treatment can be used.
This situation may be realized
in the case of the steady shear problem. 
In the later part of this paper, we only discuss the cases satisfying eq.(\ref{steady-cd}). 
Note that $\langle A(t) \rangle$ is time-independent, though $A(\Gamma(t))$ can be time-dependent
under the steady condition (\ref{steady-cd}). 
When the steady condition (\ref{steady-cd}) is satisfied, $C_{AB}(t)$ is reduced to
\begin{equation}\label{n-17}
C_{AB}(t)=\int d\Gamma \rho(\Gamma)A(\Gamma)e^{-i\cl^-_{tot} t}[B^*(\Gamma)]=\langle A B^*(-t)\rangle,
\end{equation}
where we use the translational invariance of the correlation function, {\it i.e.},  $C_{AB}(t)=\langle A(t)B^*\rangle=\langle A B^*(-t)\rangle$.
Thus, from eq.(\ref{n-17}) with the aid of eq.(\ref{steady-cd})  we obtain
\begin{equation}\label{adjointeq}
A^*(\Gamma(t))=e^{i\cl^-_{tot}t}A^*(\Gamma);\quad \frac{dA^*(\Gamma(t))}{d{t}}=i\cl^-_{tot}A^*(\Gamma(t))
\end{equation}
by the replacement of $B^*(\Gamma(t))$ by $A^*(\Gamma(t))$. 

We may arise the naive question whether it is possible to derive the generalized Langevin equation for sheared granular fluids under the assumption of eq.(\ref{steady-cd}).
The answer of this question is "yes".

The procedure to derive the generalized Langevin equation is parallel to that of the classical simple liquid\cite{zwanzig,hansen} except for the non-self adjoint properties of $i\cl_{tot}$.
The details of the derivation are given in Appendix A. 

The final expression of the generalized Langevin equation is 
\begin{equation}\label{note54}
\dot{A}(t)-i\Omega A(t)+\int_0^tds M(t-s)A(s)=R(t).
\end{equation}
Here the memory kernel $M(t)$ is given by
\begin{equation}\label{note55}
M(t)=\frac{(\bar{R},R(t))}{(A,A)} ,
\end{equation}
and $\Omega$ is
\begin{equation}\label{Omega:3.7}
i\Omega\equiv \frac{(A,\dot{A})}{(A,A)}.
\end{equation}
The random force in eq.(\ref{note54}) is 
\begin{equation}\label{R(t)R}
R(t)=e^{i\cQ\cl_{tot}\cQ t}\cQ \cl_{tot} A, \quad \bar{R}=i\cQ\cl^-_{tot} A,
\end{equation}
where we   introduce  the projection operator
\begin{equation}\label{projection}
\cP B(t)\equiv \frac{(A,B(t))}{(A,A)}A; \quad \cQ\equiv 1-\cP ,
\end{equation}
and the inner product 
\begin{equation}\label{inner-product}
(A,B(t))\equiv \langle B(t)A^*\rangle=\int d\Gamma B(t)A^*\rho(\Gamma).
\end{equation}

It is one of the most important results in this paper to obtain the generalized Langevin equation
(\ref{note54})  supplemented by eqs.(\ref{note55})-(\ref{inner-product}).  
Although nobody has discussed the generalized Langevin equation for granular fluids, the equation is useful in particular to construct a "liquid theory"
of granular fluids.

\section{Some formulae in the presence of the shear}

The treatment of the generalized Langevin equation (\ref{note54}) for sheared granular liquids is still difficult
because of the existence of $i\cl_s$, 
but the effects of shear
 can be absorbed for the description of the hydrodynamic variables.
For example, the operation of $i\cl_s$ to the local density field $n(\Vect{r},t)$ which is defined by
\begin{equation}\label{IV-1}
n(\Vect{r},t)=\sum_{j=1}^N\delta(\Vect{r}-\Vect{r}_j(t))
\end{equation}
for the position of $j-$th particle $\Vect{r}_j(t)$ is given by
\begin{eqnarray}
i\cl_sn(\Vect{r},t)&=&\dgam\sum_j y_j\frac{\partial}{\partial x_j}\sum_k\delta(\Vect{r}-\Vect{r}_k(t))=
\dgam\sum_j y_j\frac{\partial}{\partial x_j}\delta(\Vect{r}-\Vect{r}_j(t))
\nonumber\\
&=& -\dgam y\frac{\partial}{\partial x}\sum_j\delta(\Vect{r}-\Vect{r}_j(t))=-\dgam y\frac{\partial}{\partial x}n(\Vect{r},t) .
\end{eqnarray}
Therefore, the density field obeys
\begin{equation}
\frac{\partial n(\Vect{r},t)}{\partial t}+\dgam y\frac{\partial}{\partial x}n(\Vect{r},t)=i\cl n(\Vect{r},t) .
\end{equation}
Similarly, any field variable $A(\Vect{r},t)$ obeys
\begin{equation}\label{51v1}
\frac{dA(\Vect{r},t)}{dt}\equiv\frac{\partial A(\Vect{r},t)}{\partial t}+\dgam y\frac{\partial}{\partial x}A(\Vect{r},t)=i\cl A(\Vect{r},t) ,
\end{equation}
where $i\cl$ is the sum of the free motion and the inelastic collisions acting on 
the hydrodynamic variable $A(\Vect{r},t)$. 
Thus, the time evolution in sheared systems is governed by the Liouvillian  $i\cl$ 
instead of the total Liouvillian $i\cl_{tot}$. Therefore, $i\cl_{tot}$ and $i\cl^-_{tot}$ in the previous section and Appendix A can be formally replaced by
$i\cl$ and $i\cl^-$ in the later discussion, respectively.
We also note that our problem is reduced to a standard setup of field variables in the presence of 
the shear.\cite{fuchs05,onuki,miyazaki04}

It should be noted that the second term in the left hand side of eq.(\ref{51v1}) can be eliminated 
in the sheared frame
\begin{equation}\label{shear-frame}
\tilde{A}(\Vect{r}_t,\tilde{t})=A(\Vect{r},t), \quad \Vect{r}_t=\Vect{r}-\dot{\gamma}y t\Vect{e}_x,
\end{equation}
where $\tilde{t}=t$ and $\Vect{e}_x$ is the unit vector along $x-$coordinate in the experimental frame.
Indeed, $\tilde{A}(\Vect{r}_t,\tilde{t})$ obeys
\begin{equation}
\frac{\partial \tilde{A}(\Vect{r}_t,\tilde{t})}{\partial \tilde{t}}=i\cl \tilde{A}(\Vect{r}_t,\tilde{t})
\label{sheared-eq}
\end{equation}
where we use 
\begin{equation}
\partial_t=\partial_{\tilde{t}}-\dot{\gamma}y\partial_{\tilde{x}}, \quad 
\partial_y=\partial_{\tilde{y}}-\dot{\gamma}t\partial_{\tilde{x}}, \quad
\partial_{x_{\alpha}}=\partial_{x_{\alpha}} \quad {\rm for}{~} \alpha=1,3
\end{equation}
in the three dimensional case.

Let us introduce the Fourier transform
\begin{equation}
A_{\Vect{q}}(t)=\int d\Vect{r}e^{i\Vect{q}\cdot\Vect{r}}A(\Vect{r},t) ,
\quad \tilde{A}_{\Vect{q}_t}(\tilde{t})=\int d\Vect{r}_t e^{i\Vect{q}_t\cdot\Vect{r}_t}\tilde{A}(\Vect{r}_t,\tilde{t})
\end{equation}
where $\Vect{q}_t$ is the stretched wave number related to the simple wave number $\Vect{q}=(q_x,q_y,q_z)$ 
\begin{equation}\label{q(t)}
\Vect{q}_t\equiv (q_x,q_y+\dgam q_x t,q_z)
\end{equation}
in the three dimensional system. It is easy to show the equivalences of these two representations 
\begin{equation}\label{fourier-rel}
\tilde{A}_{\Vect{q}_t}(\tilde{t})=A_{\Vect{q}}(t).
\end{equation}
 We also obtain
the time evolution of the field variable $\tilde{A}_{\Vect{q}_t}(\tilde{t})$ in the sheared frame is given by
\begin{equation}\label{renorm-eq}
\frac{\partial}{\partial \tilde{t}}\tilde{A}_{\Vect{q}_t}(\tilde{t})=
i\cl_{\Vect{q}_t}\tilde{A}_{\Vect{q}_t}(\tilde{t}) ,
\end{equation}
where we use
\begin{equation}
\partial_t=\partial_{\tilde{t}}+\dot{\gamma}q_x\partial_{q_y}, \quad 
\frac{\partial}{\partial q_x}=\frac{\partial}{\partial{q_t}_x}+\dot\gamma \frac{\partial}{\partial{q_t}_y}, \quad 
\frac{\partial}{\partial q_{\alpha}}=\frac{\partial}{\partial {q_t}_{\alpha}} \quad {\rm for}{~}\alpha = 2,3.
\end{equation}
The right hand side of eq.(\ref{renorm-eq}) includes the Liouvillian $i\cl_{\Vect{q}_t}$
which is the result of the Fourier transform in the sheared frame.

Here, we note that the introduction of Fourier transform  implicitly assumes to use
the periodic boundary condition such as Lees-Edwards condition. Fortunately, it is known that rheological 
properties of dense granular flows under the shear are little affected by the choice of 
the boundary conditions.\cite{hatano07a,hatano07b} Thus, we believe that our theoretical argument in this
paper can be used even in physical situations.

In Appendix B, we summarize the relevant expression of the correlation function in the presence of the shear.
This Appendix might be useful to resolve the confusion among various expressions in literature.

\section{Generalized Langevin equation for the sheared granular liquids}

In section 3, we have derived the generalized Langevin equation, but the result is too formal
 to apply it to physical processes. Since our objective is to derive the mode-coupling theory (MCT) for sheared granular
liquids, we had better focus our attention on the density fluctuations. For this purpose we first summarize what
we calculate to derive MCT equation.

MCT is the theory to describe the time correlation function of the density fluctuations. The density of liquids
is given by eq.(\ref{IV-1}) and its Fourier component in the experimental frame becomes
\begin{equation}\label{V-1}
n_{\Vect{q}}(t)=\sum_{j=1}^Ne^{i\Vect{q}\cdot\Vect{r}_j(t)} .
\end{equation}
From the argument in Appendix B, 
the density correlation function in the experimental frame should be characterized by $F(\Vect{q},t)$ 
defined in eq.(\ref{B.5}) 
 where suffices $A=B$ for the density field in Appendix B are eliminated.
Let us use $F(\Vect{q},t)$ for later discussion, in which the explicit definition is given by
\begin{equation}
F(\Vect{q},t)\equiv\frac{1}{N}\langle n_{\Vect{q}_{-t}}(t)n_{-\Vect{q}}(0)\rangle =
\frac{1}{N}\sum_{j,k}\exp[i(\Vect{q}_{-t}\cdot\Vect{r}_j(t)-\Vect{q}\cdot\Vect{r}_k)] .
\end{equation}
The function $F(\Vect{q},t)$ is reduced to the scattering function at $t=0$
\begin{equation}
F(\Vect{q},t=0)=\frac{1}{N}\langle n_{\Vect{q}}(0)n_{-\Vect{q}}(0)\rangle\equiv S(\Vect{q}).
\end{equation}
Note that the form of the structure factor $S(\Vect{q})$ or the pair correlation function $g(\Vect{r})$ which is the Fourier transform of $S(\Vect{q})$
for granular fluids has not been established.
There are some theoretical studies\cite{lutsko01} without any external forces but no theoretical work 
in the presence of the shear. 
However, the theoretical approaches suggest that the structure factor is 
independent of the relaxation processes in granular liquids, which should be valid in our steady case (\ref{steady-cd}). 
Therefore,  we can separate the problem to
determine $F(\Vect{q},t)$ from the determination of $S(\Vect{q})$. 
We also note
that the anisotropy of the 
pair correlation function induced by the shear exists but is small in the simulations of 
sheared granular flows\cite{alam,namiko07}
as in the case of sheared colloidal particles. 
One of the characteristics, however, for granular liquids appears in $g(\Vect{r})$ where the first peak around $|\Vect{r}|=\sigma$ is higher than that in
the conventional cases.\cite{lutsko01,namiko07}

Consider the density defined in eq.(\ref{V-1}). Its time evolution is described by
\begin{equation}\label{RC17'}
\frac{d}{d t}n_{\Vect{q}_{-t}}(t)=iq_{-t}j_{\Vect{q}_{-t}}^L(t) .
\end{equation}
Here we introduce $j_{\Vect{q}_{-t}}^L(t)\equiv \hat{\Vect{q}}_{-t}\cdot\Vect{j}_{\Vect{q}_{-t}}(t)$
 with the current
\begin{equation}
\Vect{j}_{\Vect{q}_{-t}}(t)\equiv \sum_{j}\dot{\Vect{r}}_j(t)e^{i\Vect{q}_{-t}\cdot\Vect{r}_j(t)} 
\end{equation} 
and $\hat{\Vect{q}}_{-t}\equiv \Vect{q}_{-t}/q_{-t}$.

To describe the slow process of the density correlation function
 in sheared granular liquids we formally replace $A(t)$  and $i\cl_{tot}$ in section 3
 by 
\begin{equation}
\Vect{A}_{\Vect{q}_{-t}}(t)=
\left(\begin{array}{c}
\delta n_{\Vect{q}_{-t}}(t) \\
j_{\Vect{q}_{-t}}^L(t)
\end{array}\right)
\end{equation}
 and $i\cl_{\Vect{q}_{-t}}$, respectively, 
where $\delta n_{\Vect{q}_{-t}}(t)=\sum_je^{i\Vect{q}_{-t}\cdot\Vect{r}_j(t)}-(2\pi)^d\bar{n}\delta(\Vect{q}_{-t})$
 with the average density $\bar{n}$, and the longitudinal mode of
Fourier component of the current field. 

Therefore, eq.(\ref{note54}) is replaced by
\begin{equation}\label{RC25}
\frac{d\Vect{A}_{\Vect{q}_{-t}}(t)}{dt}=i\Vect{\Omega}_{\Vect{q}}\cdot\Vect{A}_{\Vect{q}_{-t}}(t)-
\int_0^{t}ds\Vect{M}_{\Vect{q}_{-(t-s)}}
(t-s)\cdot\Vect{A}_{\Vect{q}_{-s}}(s)+\Vect{R}_{\Vect{q}_{-t}}(t) ,
\end{equation}
where the memory kernel, the random force and the characteristic frequency are respectively given by
\begin{equation}\label{n-63}
\Vect{M}_{\Vect{q}_{-t}}
(t)=
(\bar{\Vect{R}}_{\Vect{q}},\Vect{R}_{\Vect{q}_{-t}}(t))(\Vect{A}_{\Vect{q}},\Vect{A}_{\Vect{q}})^{-1} ,
\end{equation} 
\begin{equation}\label{eq:5.9}
\Vect{R}_{\Vect{q}_{-t}}(t)=\exp[i\int_0^t \cQ\cl_{\Vect{q}_{-t'}}\cQ dt']i\cQ \cl_{\Vect{q}}\Vect{A}_{\Vect{q}} ,
\quad
\bar{\Vect{R}}_{\Vect{q}}=i\cQ \cl^-_{\Vect{q}}\Vect{A}_{\Vect{q}} ,
\end{equation}
and
\begin{equation}\label{66aa}
i\Vect{\Omega}_{\Vect{q}}\equiv (\Vect{A}_{\Vect{q}},\dot{\Vect{A}}_{\Vect{q}})(\Vect{A}_{\Vect{q}},\Vect{A}_{\Vect{q}})^{-1}.
\end{equation}
We also introduce the correlation matrix
\begin{equation}\label{RC27}
\Vect{C}_{\Vect{q}_{-t}}(t)\equiv \langle\Vect{A}_{\Vect{q}_{-t}}(t)
{\Vect{A}^*}_{\Vect{q}}(0)\rangle=(\Vect{A}_{\Vect{q}},\Vect{A}_{\Vect{q}_{-t}}(t)) .
\end{equation}
Using $(\Vect{A},\Vect{R}(t))=0$ we obtain
\begin{equation}\label{RC28}
\frac{d\Vect{C}_{\Vect{q}_{-t}}(t)}{dt}=i\Vect{\Omega}_{\Vect{q}}\cdot\Vect{C}_{\Vect{q}_{-t}}(t)
-\int_0^{t}ds
\Vect{M}_{\Vect{q}_{-(t-s)}}(t-s)\cdot\Vect{C}_{\Vect{q}_{-s}}(s) ,
\end{equation}
where $\Vect{C}_{\Vect{q}_{-t}}(t)$ is
\begin{equation}\label{66}
\Vect{C}_{\Vect{q}_{-t}}(t)=
\left(
\begin{array}{cc}
\langle \delta n_{-\Vect{q}}n_{\Vect{q}_{-t}}(t)\rangle & \langle \delta n_{-\Vect{q}} j_{\Vect{q}_{-t}}^L(t)\rangle \\
\langle j_{-\Vect{q}}^L \delta n_{\Vect{q}_{-t}}(t)\rangle & \langle j_{-\Vect{q}}^L j_{\Vect{q}_{-t}}^L(t)\rangle 
\end{array}
\right),
\end{equation}
which reduces to 
\begin{equation}\label{70aa}
\Vect{C}_{\Vect{q}}(0)=\left(
\begin{array}{cc}
NS(\Vect{q}) & 0 \\
0  & \frac{NT}{m}
\end{array}
\right) 
\end{equation}
at the equal time case ($t=0$), where we use the granular temperature as $T$.
Here, we should add some explanations for eq.(\ref{70aa}). The diagonal elements of (\ref{70aa}) are respectively  the definitions of
 the structure factor and the granular temperature. 
On the other hand, the off-diagonal element becomes 
$\langle j_{-\Vect{q}}^L\delta n_{\Vect{q}} \rangle=m^{-1}\sum_{i}\langle \hat{\Vect{q}}\cdot\Vect{p}_i \rangle
-m^{-1}\sum_i\langle (\hat{\Vect{q}}\cdot\Vect{p}_i)e^{-i\Vect{q}\cdot\Vect{r}_i}\delta(\Vect{q})\rangle (2\pi)^3\bar{n}$,
where $\Vect{p}_i=m(\Vect{v}_i-\dot{\gamma}y\Vect{e}_x)$ is the linear momentum of the particle $i$.
It is natural to assume that the momentum is independent of the position and the average of linear term of the momentum 
is zero in the statistical average of the uniform shear. Thus, we assume that the off-diagonal element in eq.(\ref{70aa}) 
is zero.

From eqs.(\ref{66aa}), (\ref{RC27}), (\ref{66}) and (\ref{70aa}) we also have the expression
\begin{eqnarray}
i\Vect{\Omega}_{\Vect{q}} &=&\left(
\begin{array}{cc}
\langle \delta n_{-\Vect{q}}  \frac{d}{dt}\delta n_{\Vect{q}} \rangle & 
\langle \delta n_{-\Vect{q}}  \frac{d}{dt}j_{\Vect{q}}^L\rangle \\
\langle j_{-\Vect{q}}^L\frac{d}{dt}\delta n_{\Vect{q}} \rangle
& \langle j_{-\Vect{q}}^L \frac{d}{dt}j_{\Vect{q}}^L\rangle
\end{array}
\right)
\langle \Vect{A}^*_{\Vect{q}}\Vect{A}_{\Vect{q}} \rangle^{-1}
\nonumber\\
 &=& \left(
\begin{array}{cc}
0 & i\frac{Nq T}{m} \\
i\frac{N q T}{m} 
& 0
\end{array}
\right)
\left(
\begin{array}{cc}
\frac{1}{NS(\Vect{q})} & 0 \\
0 & \frac{m}{NT}
\end{array}
\right)
=\left(
\begin{array}{cc}
0 & iq \\
i\frac{q T}{mS(\Vect{q})} & 0
\end{array}
\right) .
\label{68}
\end{eqnarray}
Here we assume that the correlation between a field variable and its time derivative is always zero. 
We also use the following identity
\begin{eqnarray}
\langle j_{\Vect{q}}^L\delta \dot{n}_{\Vect{q}}\rangle &=&
i\sum_{j,k}\langle(\hat{\Vect{q}}\cdot(\dot{\Vect{r}}_j-\dot{\gamma}y\Vect{e}_x))e^{-i\Vect{q}\cdot\Vect{r}_j}
(\Vect{q}\cdot(\dot{\Vect{r}}_k-\dot{\gamma}y \Vect{e}_x))e^{i\Vect{q}\cdot\Vect{r}_k} \rangle \nonumber \\
&=&\frac{iq}{md}\sum_i\langle m(\Vect{v}_j-\dot{\gamma}y\Vect{e}_x)^2\rangle=i\frac{NqT}{m} 
\end{eqnarray}
to obtain eq.(\ref{68}).
It should be noted that the granular temperature $T$ is proportional to $\dot\gamma^2$ in granular fluids
 under the steady shear.\cite{namiko}
 For simplicity, we assume that the temperature is uniform, which can be
realized in a periodic system in the vicinity of jamming transitions. The uniform temperature is also realized 
in the bulk region of dense granular flows under the physical boundary. 

The complex conjugate of the random force at $t=0$ is given by
\begin{equation}\label{n-68}
\bar{\Vect{R}}_{\Vect{q}}^*(0)=i\cQ^*{\cl}^-_{-\Vect{q}}\Vect{A}_{\Vect{q}}^*=i(1-\cP^*){\cl}^-_{-\Vect{q}}
\left(\begin{array}{c}
\delta n_{-\Vect{q}} \\
j_{-\Vect{q}}^L
\end{array}\right) ,
\end{equation}
where $\cP^*$ and $\cQ^*$ are the complex conjugate of the projection operators $\cP$ and $\cQ$, respectively.
From eq.(\ref{adjointeq}) with the replacement of ${\cl}^-_{tot}$ by ${\cl}^-_{\Vect{q}}$ we obtain the relation
\begin{equation}
i{\cl}^-_{-\Vect{q}} 
\left(
\begin{array}{c}
\delta n_{-\Vect{q}} \\
j_{-\Vect{q}}^L
\end{array}
\right) 
=\left(
\begin{array}{c}
\delta \dot{n}_{-\Vect{q}} \\
\dot{j}_{-\Vect{q}}^L
\end{array}
\right) .
\end{equation}
The second term in the last expression of eq.(\ref{n-68}) can be rewritten as 
\begin{eqnarray}
\cP^* i \cl^-_{-\Vect{q}}\Vect{A}_{\Vect{q}}^*&=&
\frac{(\Vect{A}_{\Vect{q}}^*,i\cl^-_{-\Vect{q}}\Vect{A}_{\Vect{q}}^*)}{(\Vect{A}_{\Vect{q}}^*,\Vect{A}_{\Vect{q}}^*)}
\Vect{A}_{\Vect{q}}^*
=\frac{(\Vect{A}_{\Vect{q}}^*,\dot{\Vect{A}}_{\Vect{q}}^*)}{(\Vect{A}_{\Vect{q}}^*,\Vect{A}_{\Vect{q}}^*)}
\Vect{A}_{\Vect{q}}^* \nonumber\\
&=& 
\left(
\begin{array}{cc}
0 & -i q \\
-i\frac{qT}{mS(\Vect{q})} & 0
\end{array}
\right)
\left(
\begin{array}{c}
\delta n_{-\Vect{q}} \\ j_{-\Vect{q}}^L
\end{array}
\right)
=
\left(
\begin{array}{c}
-iq j_{-\Vect{q}}^L
\\ -i\frac{q T}{m S(\Vect{q})}\delta n_{-\Vect{q}}
\end{array}
\right).
\end{eqnarray}
Therefore,  we obtain
\begin{equation}
\bar{\Vect{R}}_{\Vect{q}}^*=
\left(
\begin{array}{c}
0 \\
\displaystyle\frac{dj_{-\Vect{q}}}{dt}+iq\frac{T}{mS(\Vect{q})}\delta n_{-\Vect{q}}
\end{array}
\right)
=
\left(
\begin{array}{c}
0 \\ \bar{R}_{-\Vect{q}}
\end{array}
\right) .
\label{73}
\end{equation}
Similarly, $R_{\Vect{q}_{-t}}(t)$ is given by
\begin{equation}\label{RC40}
R_{\Vect{q}_{-t}}(t)=\frac{dj_{\Vect{q}_{-t}}^L(t)}{dt}-i\frac{q_{-t} T}{m S(\Vect{q}_{-t})}\delta n_{\Vect{q}_{-t}}(t),
\end{equation}
where we use eq.(\ref{eq:5.9}).

Let us look at the equation of motion term by term. First, the left hand side of eq.(\ref{RC28}) becomes
\begin{equation}
\frac{d\Vect{C}_{\Vect{q}_{-t}}(t)}{dt}
=\left(
\begin{array}{cc}
\frac{d}{dt}\langle \delta n_{-\Vect{q}} \delta n_{\Vect{q}_{-t}}(t)\rangle & 
\frac{d}{dt} \langle \delta n_{-\Vect{q}} j_{\Vect{q}_{-t}}^L(t) \rangle \\
\frac{d}{dt}\langle j_{-\Vect{q}}^L \delta n_{\Vect{q}_{-t}}(t) &
\frac{d}{dt}\langle j_{-\Vect{q}}^L j_{\Vect{q}_{-t}}^L(t)\rangle
\end{array}
\right) 
\end{equation}
in our problem by using eq.(\ref{66}).
Note that the lower left corner element becomes $(N/iq)(d^2 F(\Vect{q},t)/d t^2)$.
Second, from eqs.(\ref{66}) and (\ref{68}) we obtain
\begin{equation}
i\Vect{\Omega}_{\Vect{q}} \cdot \Vect{C}_{\Vect{q}_{-t}}(t)
=\left(
\begin{array}{cc}
0 & iq \\
i\frac{q T}{m S(\Vect{q})} & 0
\end{array}
\right)
\left(
\begin{array}{cc}
\langle \delta n_{-\Vect{q}} \delta n_{\Vect{q}_{-t}}(t)\rangle & 
\langle \delta n_{-\Vect{q}} j_{\Vect{q}_{-t}}^L(t) \rangle \\
\langle j_{-\Vect{q}}^L \delta n_{\Vect{q}_{-t}}(t) &
\langle j_{-\Vect{q}}^L j_{\Vect{q}_{-t}}^L(t)\rangle
\end{array}
\right) ,
\end{equation}
where the lower left corner term is $-(q N T/im S(\Vect{q}))F(\Vect{q},t)$.
Lastly, the memory matrix is
\begin{equation}
\Vect{M}_{\Vect{q}_{-t}}(t)
=\displaystyle\left(\begin{array}{cc}
0 & 0 \\
0 & \langle \bar{R}_{-\Vect{q}} {R}_{\Vect{q}_{-t}}(t)\rangle 
\end{array}
\right)
\left(\begin{array}{cc}
\displaystyle\frac{1}{NS(\Vect{q})} & 0 \\
0 & \displaystyle\frac{m}{NT} 
\end{array}
\right)
=\left(
\begin{array}{cc}
0 & 0 \\
0 & \displaystyle\frac{m \langle \bar{R}_{-\Vect{q}}R_{\Vect{q}_{-t}}(t)\rangle }{NT} 
\end{array}
\right) ,
\end{equation}
where we use eqs.(\ref{68}) and (\ref{73}).
Concerning the lower left corner, using eq.(\ref{RC28}) we obtain
\begin{equation}
\label{RC39}
\frac{d^2F(\Vect{q},t)}{d t^2}+
\frac{q_{-t}^2T}{mS(\Vect{q})}F(\Vect{q},t)+
\frac{m}{NT}\int_0^{t}d\tau \langle \bar{R}_{-\Vect{q}} R_{\Vect{q}_{-\tau}}(\tau) \rangle
\frac{d}{d t}F(\Vect{q},t-\tau)=0 .
\end{equation}

\section{Mode-coupling approximation}

The derivation of eq.(\ref{RC39}) is almost exact except for the assumption of the uniform temperature 
but eq.(\ref{RC39}) is not a closed equation of $F(\Vect{q},t)$. 
On the other hand, we are only interested in the slow
dynamics of $F(\Vect{q},t)$ in the vicinity of the jamming transition. 
One of the possible approaches to get a closed equation for $F(\Vect{q},t)$ is to diagonalize the linearized 
hydrodynamic equations of $\delta n_{\Vect{q}}(t)$ and $j^L_{\Vect{q}}(t)$.\cite{otsuki07,hh-mo07}
The quantitative validity of this approach for dilute granular gases has been confirmed, 
but we may not use this approach in the dense granular liquids.  
As another approach we adopt the mode-coupling approximation for granular liquids near jamming transition  in which 
the argument borrows from that for the conventional glass transition. 
We should stress that the derivation of MCT equation is not the final goal to describe the jamming transition. 
Indeed, we will need to determine $S(\Vect{q})$ and the quantitative relation between $T$ and $\dot{\gamma}$ for granular fluids.

Let us consider the term $\langle \bar{R}_{-\Vect{q}} R_{\Vect{q}_{-t}}(t) \rangle$. 
This can be rewritten as  
$\langle \bar{R}_{-\Vect{q}} e^{i\cQ \cl_{\Vect{q}_{-t}}\cQ t}R_{\Vect{q}} \rangle$ from eq.(\ref{n-63}).
This term may consist of the fast part and the slow part. 
The fast part  may be approximated by $\Gamma_{q_{-t}} \sqrt{T} \delta(t)$ with a friction constant $\Gamma_{q_{-t}}$,
since the fast friction may be proportional to the kinetic viscosity and $q_{-t}^2$ in the hydrodynamic limit.\cite{zoppi}
The slow part represents the important contribution for the slow relaxation process, which is nothing but the mode-coupling memory kernel.

(i) Let us replace the slow part of $e^{i\cQ \cl_{\Vect{q}_{-t}} \cQ t}$ by ${\cP}_2 e^{i\cl_{\Vect{q}_{-t}}t} {\cP}_2$, 
where we introduce the new projection operator acting on any function $B$
\begin{equation}\label{RC43}
{\cP}_2 B \equiv \sum_{{\Vect{q}_1}_{-t},{\Vect{q}_2}_{-t},{\Vect{q}_3}_{-t},{\Vect{q}_4}_{-t}}
A_{{\Vect{q}_1}_{-t},{\Vect{q}_2}_{-t}}
\langle A^*_{{\Vect{q}_3}_{-t},{\Vect{q}_4}_{-t}} B \rangle
\langle A^*_{{\Vect{q}_1}_{-t},{\Vect{q}_2}_{-t}}A_{{\Vect{q}_3}_{-t},{\Vect{q}_4}_{-t}} \rangle^{-1} ,
\end{equation}
where $A_{{\Vect{q}_1}_{-t},{\Vect{q}_2}_{-t}}=\delta n_{{\Vect{q}_1}_{-t}}\delta n_{{\Vect{q}_2}_{-t}}$.
The projection operator $\cP_2$ is the simple projection onto its dominant slow mode. 
We neglect the contribution from $\cQ$ as in the case of the conventional MCT.

(ii) We assume the factorization of four-point correlation into the product of two-point ones to obtain a closed
equation for the density correlation function.

Using these approximations the derivation of MCT equation is straightforward. The details of the derivation are summarized in Appendix C.

Finally, we summarize MCT equation
\begin{equation}
\frac{d^2}{d t^2}F(\Vect{q},t)+\Gamma_{q_{-t}} \sqrt{T}\frac{d}{dt}F(\Vect{q},t)+\frac{q_{-t}^2T}{mS(\Vect{q})}F(\Vect{q},t)
+T\int_0^{t}d\tau M_{\Vect{q}_{-t}}^{MC}(\tau)\frac{d F(\Vect{q},t-\tau)}{d t}=0
\label{RC54}
\end{equation}
with
\begin{equation}\label{RC55}
M_{\Vect{q}_{-t}}^{MC}(t)=\frac{n }{16\pi^3m}
\int d\Vect{k}_{-t}  |\tilde{V}^*_{\Vect{q}-\Vect{k},\Vect{k}} \tilde{V}_{\Vect{q}_{-t}-\Vect{k}_{-t},\Vect{k}_{-t}}|
 F(\Vect{k}_{-t},t)
F(\Vect{k}_{-t}-\Vect{q}_{-t},t) 
\end{equation}
for $d=3$, where $\Vect{q}_{-t}$ are connected with $\Vect{q}$ by eq.(\ref{q(t)}), where $\tilde{V}_{\Vect{q},\Vect{k}}$ is given by (\ref{RC53}) 
with the aid of the direct correlation function (\ref{C.10}).
As suggested in section 5, the difference between sheared dense granular liquids and the conventional cases of the sheared MCT equations  appears through 
(i) the forms of the structure factor and the direct correlation function, and
(ii) the granular temperature which disappears in the limit of the low shear rate.
Although
the second term in the left hand side in (\ref{RC54}) looks dominant 
in the low temperature limit or the low shear rate limit at the first glance,
all the terms are in the same order when we assume Bagnold's scaling\cite{namiko} $T\sim \dot{\gamma}^2$ and the time
is scaled by $\dot{\gamma}^{-1}$.


\section{Discussion and conclusion}

\subsection{What can we predict from MCT equation for sheared granular liquids?}

In this paper, we have demonstrated that the MCT equation can be derived in sheared dense granular liquids.
This is the first step to understand the universal feature of the dense granular liquids and the jamming transition.
Let us summarize what we can predict for sheared dense granular liquids from MCT equation (\ref{RC54}) supplemented 
with (\ref{RC55}). 

First, we expect that there is a plateau in the relaxation of $F(\Vect{q},t)$ at least for nearly elastic granular particles when the density is larger than 
a certain threshold value. Indeed, eqs.(\ref{RC54}) and (\ref{RC55}) reduce to the well-known MCT equation 
for relatively short time $t\ll \tau(\dot{\gamma},e)$ where the characteristic time (or the life time) of the plateau
$\tau(\dot{\gamma},e)$ may satisfy
$\tau(\dot{\gamma},e)\propto \dot{\gamma}^{-1}$ at least for nearly elastic cases.\cite{fuchs02} 
Then MCT equation exhibits a quasi-arrested state of particles in cages. However, this arrested state is destructed 
by the stretching of cages induced by the sheared force. Therefore, we may write
\begin{equation}\label{arrest}
F(\Vect{q},t)\simeq \Phi(T(\dot\gamma,e))\tilde{F}(\Vect{q},t/\tau(\dot{\gamma},e))
\end{equation} 
in the quasi-arrested state. 
Here $\Phi(T)$ represents the scale factor as a function of the temperature, which tends to unity in the limit of
$T\to 0$ or $\dot\gamma\to 0$.
We also note that there is no steady state in the limit of $e=1$ because of the viscous heating effect.\cite{comment}
Although
so far there was no report of the existence of  visible plateau in granular liquids\cite{douchat07}, 
we have reproduced a two-step relaxation of $F(\Vect{q},t)$ 
from the simulation of a dense and nearly elastic sheared granular liquid as explained later. 
Similar to the conventional cases\cite{barrat}
to reproduce the two-step relaxation, we need to prepare binary systems. Indeed,
there is the crystalization for mono-disperse spheres, while there is no plateau for randomly dispersed particles.
In addition, we note that the range of parameters to observe two-step relaxations seems to be narrower than the conventional cases.

Let us explain the preliminary result of our simulation briefly. The system simulated is a three-dimensional 
80:20 mixture of $N=1000$
  Lennard-Jones system in which the potential is given by
\begin{equation}
V(\Vect{r}_{\alpha\beta})=4\epsilon_{\alpha\beta}
\left[\left(\frac{\sigma_{\alpha\beta}}{r_{\alpha\beta}}\right)^{12}-
\left(\frac{\sigma_{\alpha\beta}}{r_{\alpha\beta}}\right)^{6}
\right] ,
\end{equation}
where $\alpha$ and $\beta$ refer to two speicies, and $r_{\alpha\beta}$ denotes the distance between
the particle $\alpha$ and the particle $\beta$. The particles are confined in a periodic box whose linear dimension
is 9.4$\sigma_{AA}$ under the Lees-Edwards boundary condition.
We choose $\epsilon_{AB}=1.5\epsilon_{AA}$, $\epsilon_{BB}=0.5\epsilon\/{AA}$, $\sigma_{BB}=0.88\sigma_{AA}$ and $\epsilon_{AB}=0.8\sigma_{AA}$.
We introduce the dissipative force 
$-\eta (\Vect{v}_{\alpha\beta}\cdot\Vect{r}_{\alpha\beta})\Vect{r}_{\alpha\beta}/r_{\alpha\beta}^2$
with $\Vect{r}_{\alpha\beta}=\Vect{r}_{\alpha}-\Vect{r}_{\beta}$ and 
$\Vect{v}_{\alpha\beta}=\Vect{v}_{\alpha}-\Vect{v}_{\beta}$
 for $r_{\alpha\beta}<1.12246\sigma_{\alpha\beta}$
in the equation of motion of the particle $\alpha$, where the dissipation parameter $\eta=0.001$ in the dimensionless unit.
Note that all quantities are non-dimensionalized by the particles diameter $\sigma_{AA}$, the interaction energy $\epsilon_{AA}$ and the time 
$\tau_0=(m_A\sigma_{AA}^2/\epsilon_{AA})^{1/2}$ with the mass of A particle $m_A$.
We add the shear with $\dot\gamma=0.001/\tau_0$ to the system. 
Figure 1 is the result of the mean sqaure displacement of particles 
$\langle r(t)^2\rangle \equiv \sum_i\langle (z_i(t)-z_i(0))^2 \rangle/N$, where
$z_i(t)$ is the position of $i-$th particle in the vertical direction to the sheared plane.
It is clear that the particles are in a quasi-arrested state in the middle stage of time. 
Figure 2 is the result of $F^s(q,t)\equiv  \sum_i\langle \cos (q(z_i(t)-z_i(0))\rangle/N$ 
with $q=15$ in the dimensionless unit.
We find that $F(q,t)$ has the plateau in the middle of the relaxation process. 
The details of our simulation will be reported elsewhere.

     \begin{figure}
         \centerline{\includegraphics[width=8 cm]
                                     {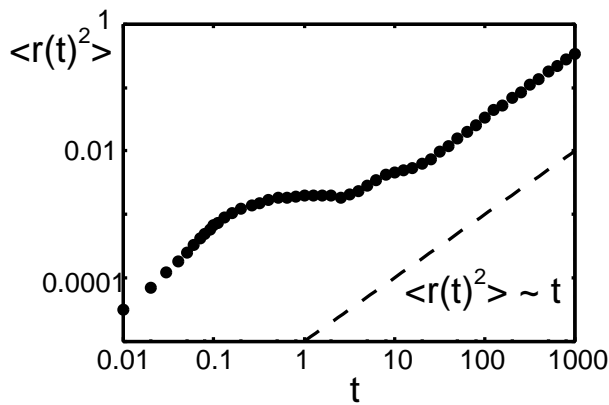}}
     \caption{The mean square displacment of particles 
$\langle r(t)^2\rangle \equiv\sum_i\langle (z_i(t)-z_i(0))^2\rangle/N$ as a function of time .}
     \label{fig:1}
     \end{figure}
 \begin{figure}
         \centerline{\includegraphics[width=8 cm]
                                     {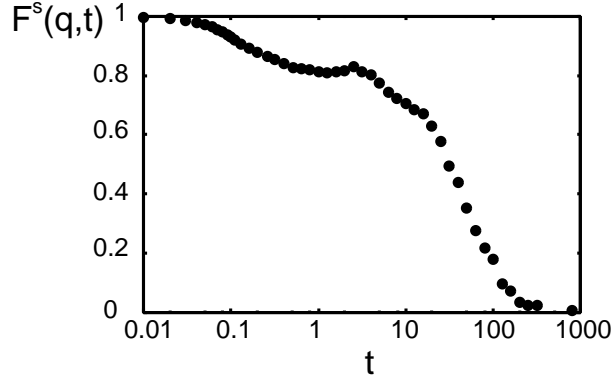}}
     \caption{The relaxation of the correlation function $F^s(q,t)=\sum_i\langle \cos (q(z_i(t)-z_i(0))\rangle/N$.}
     \label{fig:2}
     \end{figure}


Second, it is difficult to describe the jamming transition based on the hard-core model which we use in this paper. 
Let us demonstrate this difficulty as follows. 
We expect that the plateau $\tau(\dot{\gamma},e)\propto \dot{\gamma}^{-1}$ becomes
long as $\dot\gamma$ decreases. Eventually, the plateau becomes infinitely long in the limit of $\dot\gamma\to 0$.
This is the reflection of no motion of particles in the limit of $\dot\gamma\to 0$.
When we adopt the factorization approximation as in the framework of MCT, the shear stress $\sigma_{xy}$
for sheared granular liquids 
can be represented by a function of 
$S(\Vect{q})$, $S(\Vect{q}_t)$ and $F(\Vect{q},t)$  as\cite{fuchs02}
\begin{equation}\label{fuchs-cates}
\sigma_{xy}=\frac{T\dot{\gamma}}{60\pi^2}\int_0^{\infty}dt\int_0^{\infty}dk k^4\frac{S'(k)S'(k_t)}{S(k_t)^2}\left(\frac{F(k,t)}{S(k)}\right)^2
\end{equation}
where $S'(k)\equiv \partial S(k)/\partial k$, and 
we ignore anisotropy of the structure factor and the density correlation function. 
To derive eq.(\ref{fuchs-cates}) we use the decoupling approximation and 
the Green-Kubo formula which may be suspicious in granular liquids. 
However, the decoupling approximation is consistent with MCT and
the Green-Kubo formula
gives, at least, a good approximate expression for the nearly elastic granular liquids. Thus, the expression (\ref{fuchs-cates}) should be valid
for the nearly elastic granular liquids. 
If we assume that there is no relaxation of $F(k,t)$ in a quasi-arrested state,  eq.(\ref{fuchs-cates}) may be replaced by
\begin{eqnarray}
\sigma_{xy} &\simeq& \frac{T\dot{\gamma}}{60\pi^2} 
F(k,\tau(\dot{\gamma},e))^2\int_0^{\tau(\dot{\gamma},e)}dt\int_0^{\infty} dk k^4\frac{S'(k)S'(k)}{S(k)^2} \nonumber \\
&\propto & 
T\dot{\gamma}F(k,\tau(\dot{\gamma},e))^2\tau(\dot{\gamma},e)\propto T,
\label{eq7.3}
\end{eqnarray}
where we replace the integrand by that for $t\ll \tau(\dot{\gamma},e)\sim\dot{\gamma}^{-1}$.
From eq.(\ref{eq7.3})  the viscosity $\eta$ can be evaluated as $\eta\sim T|\dot{\gamma}|^{-1}$  as in the case of colloidal suspensions near the glass transition.
\cite{miyazaki02,fuchs02, miyazaki04,fuchs05} 
This does not mean that the sheared granular liquid displays the shear thinning property, 
but this result is consistent with Bagnold's scaling $\eta\sim |\dot{\gamma}|$ or $T\sim \dot{\gamma}^2$  for hard-core granular fluids\cite{namiko}
which is one of the shear thickenning relations. Indeed, if we assume that the continuity equation for the energy is still relevant, the uniform steady  granular liquids obey
$\eta \dot{\gamma}^2\sim T^{3/2}$. Therefore,  we obtain the relation
\begin{equation}\label{bagnold}
T\sim \dot{\gamma}^2
\end{equation}
which is nothing but Bagnold's scaling.

From eqs.(\ref{eq7.3}) and (\ref{bagnold})  we may obtain
\begin{equation}\label{sigma}
\sigma_{xy} \sim T\sim  \dot{\gamma}^2.
\end{equation}
We should note that Bagnold's scaling (\ref{bagnold}) is violated in the vicinity of the jamming transition as
 $T\sim \dot{\gamma}^{\beta}$ with $\beta\simeq 1.1$.\cite{hatano08}
We also note that the jamming may be the continuous transition\cite{hatano08,olsson}, 
which is different from the sheared dynamic yield stress at a constant temperature.

Equations (\ref{sigma}) may imply the following two things: 
(i)  The stress can be indepent of the shear rate when we keep a constant temperature with changing the restitution constant $e$.
(ii) However, the stress becomes zero in the limit of $\dot{\gamma}\to 0$. 
Thus,  we may conclude
that our MCT for hard-spheres is invalid to describe the jamming transition.
This conclusion is reasonable, because the jamming transition is originally defined by the point of non-zero bulk and shear moduli at zero temperature
without kinetic energy\cite{O'hern}.  


\subsection{Future problems}

In this paper, we only present the formal derivation of MCT equation and give some qualitative predictions. 
To know the quantitative details of the behaviors of dense
granular liquids, we need to know the explicit expression of $S(\Vect{q})$ and the relation
between the granular temperature and the shear rate. As long as we know, we do not have any satisfactory  theory
to describe sheared dense granular liquids. This will be our important subject to the future.

We should note that 
our MCT is based on the mean field description in which the system is almost uniform. 
The heterogeneity may be important in the actual jamming transitions,
but this heterogeneity may disappear if we only observe two-point correlation functions 
as in the case of the glass transition.  We believe that we will need to
formulate the model including higher correlation to describe such heterogeneity.

We also note that our starting equation is the Liouville equation for hard spherical particles. The system of inelastic hard spheres may cause the inelastic collapse
in which the collision frequency is divergent when the density becomes extremely high
in the vicinity of the jamming transition.\cite{jaeger96} 
We also note that the jamming is caused by the simultaneous contacts among particles\cite{O'hern} which cannot be described by hard-core models.
Although we obtain the formal expression for MCT equation, we may
have to extend our formulation to soft spheres to describe the jammed state, which was not obtained previously.

In this paper, we adopt the formal derivation based on the projection operator formalism. 
The advantage of this method is to get the exact expression in the middle of calculation, in which 
equation (\ref{RC39}) should be exact. On the other hand, the procedure to obtain MCT equation 
in section 6 contains some uncontrolled approximations. Therefore, it is difficult to systematically improve the decoupling
approximation used to derive MCT equation. There are several efforts to overcome such difficulties in conventional
glass transitions\cite{miyazaki05,abl,kim07,nishino} based on the field theoretic approaches.
 Such methods may  need to improve our treatment
presented here.

\subsection{Conclusion}

We summarize what we have carried out in this paper. (i) We have derived the generalized Langevin equation (\ref{note54}) with (\ref{note55}) for
granular fluids under the steady condition (\ref{steady-cd}). This equation is formal and the derivation is exact.
(ii) We apply the generalized Langevin equation to the sheared case in which the stretching of 
the wave number in Liouvillian is enough to describe the sheared system  in eq.(\ref{renorm-eq}). 
This equation is also believed to be exact. (iii) We derive the equation for the density correlation function 
$F(\Vect{q},t)$
as in eq.(\ref{RC39}). 
The equation is almost exact under the assumption that the granular temperature is uniform.
(iv) We derive MCT equation (\ref{RC54}) with (\ref{RC55}) in which we include some uncontrolled approximations
as we have used to derive the MCT equation for conventional glassy systems.
(v)  We suggest the existence of a plateau of $F(\Vect{q},t)$ for the dense granular liquids.
(vi)  We also indicate that hard-core models are insufficient to describe the jamming transition.
It should be noted that the MCT equation includes the static structure factor of sheared granular materials 
which should be different from the conventional cases.

\section*{Acknowledgements}

The authors thank T. Hatano for the fruitful discussions. 
The authors also appreciate H. Wada, B. Kim and K. Miyazaki for their critical reading of this manuscript and their useful comments.
This work is partially supported by
Ministry of Education, Culture, Sports, Sciences and Technology (MEXT) Japan (Grant No. 18540371),
and the Grant-in-Aid for the 21st century COE "Center for Division and Universality in Physics" in MEXT, Japan.
One of the authors (MO) thanks the Yukawa Foundation for its financial support.

 \appendix
\section{The derivation of the generalized Langevin equation (\ref{note54})}

In this Appendix we demonstrate how to derive the generalized Langevin equation (\ref{note54}).
The procedure is parallel to that for the simple liquids.\cite{zwanzig,hansen}

Let us introduce
\begin{equation}
Y(t)\equiv \frac{(A,A(t))}{(A,A)} .
\end{equation}
With the aid of the Laplace transform
\begin{equation}
\tilde{A}(z)\equiv \int_0^{\infty}dte^{izt}A(t),
\end{equation}
eq.(\ref{Liouville}) is reduced to
\begin{equation}\label{note26}
\tilde{A}(z)=(z+\cl_{tot})^{-1}iA(0)
\end{equation}
Operating $\cP$ and $\cQ$ to eq.(\ref{note26})  we obtain
\begin{eqnarray}
z \cP \tilde{A}(z)+\cP \cl_{tot}\cP\tilde{A}(z)+\cP \cl_{tot}\cQ \tilde{A}(z) &=& i A \label{note28} \\
z \cQ \tilde{A}(z)+\cQ \cl_{tot} \cP \tilde{A}(z)+\cQ \cl_{tot} \cQ\tilde{A}(z) &=& 0 \label{note29} .
\end{eqnarray}
Substituting eq.(\ref{note29}) or the equivalent form $\cQ \tilde{A}(z)=-(z+\cQ\cl_{tot}\cQ)^{-1}\cQ \cl_{tot}\cP \tilde{A}(z)$ into eq.(\ref{note28}) we obtain
\begin{equation}\label{note33a}
z \cP\tilde{A}(z)+\cP \cl_{tot} \cP\tilde{A}(z)-\cP\cl_{tot}(z+\cQ\cl_{tot}\cQ)^{-1}\cQ \cl_{tot}\cP\tilde{A}(z)=i A .
\end{equation}
Using the relation $(A,\cP B)=(A,B)$ the inner product of eq.(\ref{note33a}) with $A$ becomes
\begin{equation}\label{note33}
z(A,\tilde{A}(z))+(A,\cl_{tot}\cP\tilde{A}(z))-(A,\cl_{tot}(z+\cQ\cl_{tot}\cQ)^{-1}\cQ\cl_{tot}\cP \tilde{A}(z))=i(A,A) .
\end{equation}
From the relation
\begin{eqnarray}
(A,i\cl_{tot}B)&=&\langle (i\cl_{tot}B)A^*\rangle =\int d\Gamma (i\cl_{tot} B)A^*\rho(\Gamma)=-\int d\Gamma (i\cl^-_{tot} A^*)B\rho(\Gamma) \nonumber\\
&=& -\langle (i\cl^-_{tot} A^*)B\rangle =-(i\cl^-_{tot}A,B) 
\end{eqnarray}
eq.(\ref{note33}) is reduced to
\begin{equation}\label{note37}
z \tilde{Y}(z)+\frac{(A,\cl_{tot}\cP \tilde{A}(z))}{(A,A)}+
\frac{({\cl}^-_{tot}A,(z+\cQ\cl_{tot}\cQ)^{-1}\cQ\cl_{tot}\cP \tilde{A}(z))}{(A,A)}=i ,
\end{equation}
where $\tilde{Y}(z)$ is the Laplace transform of $Y(t)$.

Here the second term in the left hand side of eq.(\ref{note37}) can be rewritten as
$\Omega \tilde{Y}(z)$
with $\dot{A}=dA/dt$ at $t=0$. To derive $\Omega$ in eq.(\ref{Omega:3.7}) we use the relation
\begin{equation}
\frac{(A,\cl_{tot}\cP \tilde{A}(z))}{(A,A)}=
\frac{(A,\cl_{tot}A)}{(A,A)}\frac{(A,\tilde{A}(z))}{(A,A)}
=\frac{1}{i}\frac{(A,\dot{A})}{(A,A)}\tilde{Y}(z) .
\end{equation}
On the other hand, the numerator of the third term in the left hand side of (\ref{note37}) 
can be written as
\begin{eqnarray}
({\cl}^-_{tot}A,(z+\cQ\cl_{tot}\cQ)^{-1}\cQ\cl_{tot}\cP \tilde{A}(z))
&=&
({\cl}^-_{tot}A,\cQ(z+\cQ\cl_{tot}\cQ)^{-1}\cQ\cl_{tot}A)\frac{(A,\tilde{A}(z))}{(A,A)}
\nonumber\\
&=&
({\cl}^-_{tot}A,\cQ(z+\cQ\cl_{tot}\cQ)^{-1}\cQ\cl_{tot}A)\tilde{Y}(z)
\nonumber\\
&=&(Q{\cl}^-_{tot}A,(z+\cQ \cl_{tot}\cQ)^{-1}\cQ \cl_{tot}A)\tilde{Y}(z)
\nonumber\\
&=& -(\bar{R},(z+\cQ \cl_{tot} \cQ)^{-1} R)\tilde{Y}(z) ,
\end{eqnarray}
where $\bar{R}$ is defined in eq.(\ref{R(t)R}) and
\begin{equation}\label{note42}
R\equiv i\cQ \cl_{tot}A.
, \quad \bar{R}=i\cQ {\cl}^-_{tot}A .
\end{equation}
Here we use the relations
\begin{equation}
(\cQ B,C)=((1-\cP)B,C)=(B,C)-\frac{(A,B)}{(A,A)}(A,C)
\end{equation}
and
\begin{equation}
(B,\cQ C)=(B,(1-\cP)C)=(B,C)-\frac{(B,A)}{(A,A)}(A,C)
\end{equation}
with $(A,B)=(B,A)$ for any real functions $A$, $B$ and $C$. (For complex functions, 
the relation $(B,A)=(A^*,B^*)$ should be considered).

Introducing the memory kernel in the Laplace form
\begin{equation}
\tilde{M}(z)\equiv i(\bar{R},(z+\cQ \cl_{tot} \cQ)^{-1}R)(A,A)^{-1} ,
\end{equation}
eq.(\ref{note37}) can be rewritten as 
\begin{equation}\label{note44}
-i (z+\Omega)\tilde{Y}(z)+\tilde{M}(z)\tilde{Y}(z)=1 .
\end{equation}
Therefore, we obtain 
\begin{equation}\label{note45}
\dot{Y}(t)-i\Omega Y(t)+\int_0^tdsM(t-s)Y(s)=0
\end{equation}
in terms of the inverse Laplace transform, where $M(t)$ is the memory kernel.

Equation (\ref{note45}) describes the time evolution of the correlation function.
On the other hand, the time evolution of $A(t)$ is affected by the fluctuating force.
Indeed
\begin{equation}
\hat{A}^L(z)\equiv \cQ \tilde{A}(z)
\end{equation}
obeys
\begin{eqnarray}
(z+\cQ \cl_{tot}\cQ)\hat{A}^L(z)&=& -\cQ \cl_{tot}\cP \tilde{A}^L(z)=-\cQ\cl_{tot}\tilde{Y}(z)A
\nonumber\\
&=&-\frac{1}{i}R\tilde{Y}(z) ,
\end{eqnarray}
where we use the definition of $\tilde{Y}(z)$, eqs.(\ref{note29}) and (\ref{note42}).
Thus, we obtain
\begin{equation}
\tilde{A}^L(z)=\tilde{Y}(z)\tilde{R}(z) ,
\label{note48}
\end{equation}
where $\tilde{R}(z)=i(z+\cQ \cl_{tot}\cQ)^{-1}R$. Substituting eq.(\ref{note44}) into eq.(\ref{note48})
we obtain
\begin{equation}
(-iz-i\Omega+\tilde{M}(z))\tilde{A}^L(z)=\tilde{R}(z) .
\end{equation}
Therefore we obtain
\begin{equation}
\dot{\hat{A}}(t)-i\Omega \hat{A}(t)+\int_0^tdsM(t-s)\hat{A}(s)=R(t)
\end{equation}
with
\begin{equation}
\hat{A}(t)\equiv \cQ A(t),\quad R(t)=\exp[i\cQ \cl_{tot}\cQ t]R .
\end{equation}
Since $A(t)$ satisfies $A(t)=Y A(t)+\hat{A}(t)$ we obtain (\ref{note54}).

\section{The expression of the correlation function in the presence of the shear}

Let us summarize the correlation function in the presence of the shear. The correlation function should satisfy
the translational invariance condition
\begin{equation}
\langle \tilde{A}(\Vect{r}_t+\Vect{a},\tilde{t})\tilde{B}(\Vect{r}'+\Vect{a},0)\rangle
=\langle \tilde{A}(\Vect{r}_t,\tilde{t})\tilde{B}(\Vect{r}',0)\rangle
\end{equation}
where $\Vect{a}$ is the shift vector acting on all particles in the sheared frame.
The Fourier transform of both sides of this equation leads to
\begin{equation}
\tilde{C}_{A_{\Vect{q}_t}B_{\Vect{q}'}}(t)\equiv \langle \tilde{A}_{\Vect{q}_t}(\tilde{t})\tilde{B}_{\Vect{q}'}(0)\rangle
=e^{-i(\Vect{q}_t+\Vect{q}')\cdot\Vect{a}}\langle \tilde{A}_{\Vect{q}_t}(\tilde{t})\tilde{B}_{\Vect{q}'}(0)\rangle .
\end{equation}
Thus, we obtain
\begin{equation}
\tilde{C}_{A_{\Vect{q}_t}B_{\Vect{q}'}}(t)=\langle \tilde{A}_{\Vect{q}_t}(\tilde{t})\tilde{B}_{\Vect{q}'}(0)\rangle \delta_{\Vect{q}_t,-\Vect{q}'}
\end{equation}
in the sheared frame.
From the relation (\ref{fourier-rel}) we can write
\begin{equation}\label{B.4}
C_{A_{\Vect{q}_t}B_{-\Vect{q}'}}(t)=\langle A_{\Vect{q}}(t)B_{-\Vect{q}'}(0)\rangle
=\delta_{\Vect{q}_t,\Vect{q}'}\hat{F}_{AB}(\Vect{q},t)
\end{equation} 
where $\hat{F}_{AB}(\Vect{q},t)\equiv \langle A_{\Vect{q}}(t)B_{-\Vect{q}_t}(0)\rangle$ in the experimental frame.
This relation is obtained by Fuchs and Cates.\cite{fuchs05} This relation can be rewritten as
\begin{equation}\label{B.5}
C_{A_{\Vect{q}'},B_{-\Vect{q}}}(t)=\delta_{\Vect{q}',\Vect{q}_{-t}}F_{AB}(\Vect{q},t)
\end{equation}
where $F_{AB}(\Vect{q},t)=\hat{F}_{AB}(\Vect{q}_{-t},t)=\langle A_{\Vect{q}_{-t}}(t)B_{-\Vect{q}}(0)\rangle
=\langle A_{\Vect{q}_t}(t)B_{\Vect{q}}^*(0)\rangle$, which is
obtained by Miyazaki {\it et al}.\cite{miyazaki04}

\section{The derivation of MCT equation}

Let us explain the details of the derivation of MCT equation in this Appendix.

From the two approximations mentioned in section 6 we get
\begin{equation}
\cP_2 R_{\Vect{q}_{-t}}(t)=\sum_{{{\Vect{q}_1}_{-t}},{\Vect{q}_2}_{-t}}V_{\Vect{q}_{-t}}({\Vect{q}_1}_{-t},{{\Vect{q}_2}_{-t}})
\delta n_{{\Vect{q}_1}_{-t}}(t)\delta n_{{\Vect{q}_2}_{-t}}(t) ,
\label{RC44}
\end{equation}
where
\begin{equation}\label{RC45}
V_{\Vect{q}_{-t}}({\Vect{q}_1}_{-t},{\Vect{q}_2}_{-t})=\sum_{{\Vect{q}_3}_{-t},{\Vect{q}_4}_{-t}}
\frac{
\langle \delta n_{{\Vect{q}_1}_{-t}}(t)\delta n_{{\Vect{q}_2}_{-t}}(t)R_{\Vect{q}_{-t}}(t) \rangle
}{
\langle \delta n_{{\Vect{q}_1}_{-t}}(t)\delta n_{{\Vect{q}_2}_{-t}}(t)\delta n_{{\Vect{q}_3}_{-t}}(t)
\delta n_{{\Vect{q}_4}_{-t}}(t) \rangle
} .
\end{equation}
The product of the four density fields in the denominator of eq.(\ref{RC45}) may be factorized into the products of two structure factors. 
The numerator of eq.(\ref{RC45}) is 
\begin{eqnarray}\label{n-84}
&\langle& \delta n_{\Vect{k}_{-t}}(t)\delta n_{\Vect{k}_{-t}-\Vect{q}_{-t}}(t) R_{\Vect{q}_{-t}}(t) \rangle \nonumber \\
&=& \langle \delta n_{\Vect{k}_{-t}}(t)\delta n_{\Vect{k}_{-t}-\Vect{q}_{-t}}(t)\frac{dj_{\Vect{q}_{-t}}^L(t)}{dt}\rangle
\nonumber\\
& &-i \frac{q_{-t} T}{m S(q_{-t})}\langle \delta n_{-\Vect{k}_{-t}}(t)\delta n_{\Vect{k}_{-t}-\Vect{q}_{-t}}(t)\delta n_{\Vect{q}_{-t}}(t) \rangle .
\end{eqnarray}
The first term of the above equation is
\begin{equation}\label{RC46}
\langle \delta n_{\Vect{k}_{-t}}\delta n_{\Vect{k}_{-t}-\Vect{q}_{-t}}\frac{dj_{\Vect{q}_{-t}}^L}{dt}\rangle
=-\langle \frac{d}{dt}\delta n_{-\Vect{k}_{-t}}\delta n_{\Vect{k}_{-t}-\Vect{q}_{-t}}j_{\Vect{q}_{-t}}^L\rangle -
\langle n_{-\Vect{k}_{-t}}(\frac{d}{dt}\delta n_{\Vect{k}_{-t}-\Vect{q}_{-t}})j_{\Vect{q}_{-t}}^L \rangle .
\end{equation}
Let us calculate the first term of the right hand side of eq.(\ref{RC46}).
\begin{eqnarray}
\langle \frac{d}{dt}(\delta n_{\Vect{k}_{-t}}(t)) \delta n_{\Vect{k}_{-t}-\Vect{q}_{-t}}
(t) j_{\Vect{q}_{-t}}^L(t) \rangle
&=& 
-i\langle \sum_j (\Vect{k}_{-t}\cdot\dot{\Vect{r}}_j(t))e^{-i \Vect{k}_{-t}\cdot\Vect{r}_j(t)} 
\sum_k e^{i(\Vect{k}_{-t}-\Vect{q}_{-t})\cdot\Vect{r}_k(t)} \nonumber\\
& &\times 
\sum_l (\hat{\Vect{q}}_{-t}\cdot\dot{\Vect{r}}_l(t))e^{i\Vect{q}_{-t}\cdot\Vect{r}_l(t)} \rangle \nonumber \\
&=&
-i\frac{T}{m}(\Vect{k}_{-t}\cdot\hat{\Vect{q}}_{-t}) \sum_{j,k}\langle e^{i(\Vect{k}_{-t}-\Vect{q}_{-t})\cdot\Vect{r}_j(t)}
 e^{i(\Vect{q}_{-t}-\Vect{k}_{-t})\cdot\Vect{r}_k(t)}\rangle
\nonumber\\
&=& -i(\Vect{k}_{-t}\cdot\hat{\Vect{q}}_{-t} )\frac{T}{m} NS(\Vect{k}_{-t}-\Vect{q}_{-t}) ,
\end{eqnarray}
where we use $\langle \dot{r}_{j,\alpha} \dot{r}_{l,\beta} \rangle=\delta_{jl}\delta_{\alpha\beta}T/m$.
The other term similarly gives
\begin{equation}\label{RC49}
-\langle \delta n_{-\Vect{k}_{-t}}(t)\frac{d}{dt}(\delta n_{\Vect{k}_{-t}-\Vect{q}_{-t}}(t) ) 
j_{\Vect{q}_{-t}}^L(t)\rangle
=i(\Vect{\hat{q}}_{-t}\cdot(\Vect{q}_{-t}-\Vect{k}_{-t}))\frac{T}{m}NS(\Vect{k}_{-t}) .
\end{equation}
The last term of eq.(\ref{n-84}) is hard to compute directly. 
When we adopt trhe decoupling approximation or Kirkwood approximation, we obtain\cite{goetz,green}
\begin{equation}
\label{RC50}
\langle \delta n_{-\Vect{k}_{-t}}(t)\delta n_{\Vect{k}_{-t}-\Vect{q}_{-t}}(t)\delta n_{\Vect{q}_{-t}}(t)
 \rangle
\simeq NS(\Vect{k}_{-t}) S(\Vect{q}_{-t}) S(\Vect{k}_{-t}-\Vect{q}_{-t}) .
\end{equation}
Since $\Vect{q}_{-t}$ satisfies the periodic boundary condition, we expect that the vertex function 
$V_{\Vect{q}_{-t}}({\Vect{k}_1}_{-t},{\Vect{k}_2}_{-t})$ is the function
of the difference of the wave vectors $\Vect{k}_{-t}\equiv {\Vect{k}_1}_{-t}-{\Vect{k}_2}_{-t}$. Thus, we can write
\begin{equation}
V_{\Vect{q}_{-t}}({\Vect{k}_1}_{-t},{\Vect{k}_2}_{-t})=V_{\Vect{k}_{-t},\Vect{q}_{-t}-\Vect{k}_{-t}} .
\end{equation}
Subsitituting (\ref{RC46})-(\ref{RC50}) into (\ref{RC45}) with taking  the summation over $\Vect{k}_{-t}$, we obtain
\begin{eqnarray}
V_{\Vect{k}_{-t},\Vect{q}_{-t}-\Vect{k}_{-t}}&=&
\frac{i T}{2m N}\left\{\frac{\hat{\Vect{q}}_{-t}\cdot\Vect{k}_{-t}}{S(\Vect{k}_{-t})}+\frac{\hat{\Vect{q}}_{-t}\cdot(\Vect{q}_{-t}-
\Vect{k}_{-t})}{S(\Vect{k}_{-t}-\Vect{q}_{-t})}
-(\Vect{q}_{-t}\cdot\hat{\Vect{q}}_{-t} )\right\} \nonumber\\
&=&
\frac{i n T}{2m N}\left\{(\hat{\Vect{q}}_{-t}\cdot\Vect{k}_{-t})c(\Vect{k}_{-t})+\hat{\Vect{q}}_{-t}
\cdot(\Vect{q}_{-t}-\Vect{k}_{-t})c(\Vect{k}_{-t}-\Vect{q}_{-t}) \right\} ,
\end{eqnarray}
where we introduce the direct correlation function $c(\Vect{k})$\cite{hansen} as
\begin{equation}\label{C.10}
c(\Vect{k})=\frac{1}{\bar{n}}(1-S(\Vect{k})^{-1}).
\end{equation}
Similarly, we obtain
\begin{equation}
\cP_2 \bar{R}_{-\Vect{q}}=
\sum_{\Vect{k}}V^*_{\Vect{k},\Vect{q}-\Vect{k}}\delta n_{-\Vect{k}}\delta n_{\Vect{k}-\Vect{q}} ,
\end{equation}
where
\begin{equation}
V_{\Vect{k},\Vect{q}-\Vect{k}}=\frac{i n T}{2m N}\{ 
(\hat{\Vect{q}}\cdot\Vect{k})c(\Vect{k})+\hat{\Vect{q}}\cdot(\Vect{q}-\Vect{k})c(\Vect{k}-\Vect{q})\} .
\end{equation}
Therefore,  we may obtain
\begin{eqnarray}\label{RC52}
\langle (\cP_2 \bar{R}_{-\Vect{q}})(\cP_2R_{\Vect{q}_{-t}})\rangle
&\simeq&
\sum_{\Vect{k}_{-t},\Vect{k}'}|V^*_{\Vect{k}',\Vect{q}-\Vect{k}'}V_{\Vect{k}_{-t},\Vect{q}_{-t}-\Vect{k}_{-t}}|
\langle \delta n_{-\Vect{k}'}\delta n_{\Vect{k}'-\Vect{q}}
\delta n_{\Vect{k}_{-t}}(t)\delta n_{\Vect{q}_{-t}-\Vect{k}_{-t}}(t)\rangle \nonumber\\
&\simeq& \sum_{\Vect{k}_{-t},\Vect{k}'}|V^*_{\Vect{k}',\Vect{q}-\Vect{k}'}V_{\Vect{k}_{-t},\Vect{q}_{-t}-\Vect{k}_{-t}}
|N^2F(\Vect{k}_{-t},t)F(\Vect{q}_{-t}-\Vect{k}_{-t},t) \nonumber\\
& & \times 
(\delta_{\Vect{k}_{t=0},\Vect{k}'}+\delta_{\Vect{k}_{t=0}'-\Vect{q},\Vect{k}}) \nonumber\\
&=& \frac{n^2T^2}{2m^2}\sum_{\Vect{k}}
|\tilde{V}^*_{\Vect{q}-\Vect{k},\Vect{k}}\tilde{V}_{\Vect{q}_{-t}-\Vect{k}_{-t},\Vect{k}_{-t}}|
 F(\Vect{k}_{-t},t)F(\Vect{q}_{-t}-\Vect{k}_{-t},t) ,
\end{eqnarray}
where
\begin{equation}\label{RC53}
\tilde{V}_{\Vect{q}-\Vect{k},\Vect{k}}=
\{(\hat{\Vect{q}}\cdot\Vect{k})c(\Vect{k})+\hat{\Vect{q}}\cdot(\Vect{q}-\Vect{k})c(\Vect{q}-\Vect{k})\} .
\end{equation}
Using the equations obtained in this Appendix we obtain the final expression of MCT equation (\ref{RC54}) with (\ref{RC55}).


%

\end{document}